\documentclass[preprint,prd,showpacs]{revtex4} 
 
\begin{document} 
\input{epsf}


\title{An Estimate of the Vibrational Frequencies of Spherical Virus Particles}

\author{L.H. Ford} 
 \email[Email: ]{ford@cosmos.phy.tufts.edu} 
 \affiliation{Department of Physics and Astronomy\\ 
         Tufts University, Medford, MA 02155}

\begin{abstract} 
The possible normal modes of vibration of a nearly spherical virus 
particle are discussed. Two simple models for the particle are treated,
a liquid drop model and an elastic sphere model. Some estimates for the 
lowest vibrational frequency are given for each model. It is concluded 
that this frequency is likely to be of the order of a few GHz for particles
with a radius of the order of $50 nm$.  
\end{abstract}

\maketitle 
 
\baselineskip=13pt 

\section{Introduction}

Virus particles (virions) come in a variety of sizes and shapes. However,
approximately spherical shapes with diameters in the range between $50nm$ 
and $100nm$ are especially common. Many nearly spherical viruses are revealed 
by X-ray crystallography  to have icosahedral symmetry. 
 A typical virus particle contains genetic
material, RNA or DNA, surrounded by a protein coat (capsid). 
Such an object should
have reasonably distinct vibrational frequencies, the study of which
 may be of interest.
Excitation of these vibrations could have applications in either the
diagnosis or treatment of viral diseases. To this author's knowledge,
the sole discussion of these vibrational modes in the literature is that
of Babincov{\'a} {\it et al}~\cite{Bab}. These authors discuss the conjecture
that ultrasound in the GHz range could be resonantly absorbed by HIV virus
particles, leading to their destruction. 
Cooper {\it et al}~\cite{Cooper} have recently reported the detection of
viruses by acoustic oscillations. However, the process of ``rupture event
scanning'', which these authors report, involves the separation of a virus
particle from antibodies by ultrasound. This is distinct from the 
excitation of the vibrational
modes of the virus particle itself, and occurs at much lower frequencies.

There have also been some experimental studies of ultrasonic absorption
by empty viral capsids~\cite{Cerf79,Michels85}. These experiments reveal
an enhanced absorption in the MHz range as proteins reassemble into a capsid,
but do not find a resonant peak in this frequency range. 
Witz and Brown~\cite{WB01} 
have emphasized that these and other results show that viral capsids are
flexible and change size or shape in response to vibrations or to changes 
in temperature or pH.

The purpose of the present paper is to provide some  estimates of the
lowest vibrational frequencies of a spherical virus particle. The simplest
estimate is to take this frequency to be of the order of a characteristic
speed of sound divided by the size of the virus particle. This is the estimate
used in Ref.~\cite{Bab}. For the purpose of giving a more accurate estimate,
we will examine two models, which treat the particle (1) as a liquid drop
and (2) as a uniform elastic sphere. Similar models have been used by 
Bulatov {\it et al}~\cite{Bulatov} to estimate the vibrational frequencies
of nanoclusters.

\section{A Liquid Drop Model}

Consider a sphere of radius $a$ filled with a nonviscous liquid with surface
tension $\gamma$ and mass density $\rho$. The lowest vibrational mode of 
this sphere will be a quadrupole mode with frequency~\cite{Chandra}
\begin{equation}
\nu = \frac{1}{\pi} \, \sqrt{ \frac{2 \gamma}{\rho a^3} } \, ,
\end{equation}
which can be written as
\begin{equation}
\nu = 3.4 \times 10^8 Hz \,\left(\frac{50 nm}{a}\right)^\frac{3}{2}\,
\left(\frac{\rho_W}{\rho}\right)^\frac{1}{2}\,
\left(\frac{\gamma}{\gamma_W}\right)^\frac{1}{2}\, , \label{eq:nu_drop}
\end{equation}
where $\rho_W = 10^3 kg/m^3$ and $\gamma_W = 0.073 N m$ are the mass density 
and surface tension for water, respectively. The surface tension and
mass density, along with the lowest vibrational frequency derived from
Eq.~(\ref{eq:nu_drop}) for $a = 50 nm$, are given in Table 1 for several 
liquids.

\begin{table} 
\begin{tabular}{|c||c|c|c|} \hline\hline
Liquid & $\gamma/\gamma_W$  & $\rho/\rho_W$  &  $\nu/(10^8 Hz)$ \\ \hline
Benzene  & 0.397  & 0.88  & 2.3  \\
Diethylene glycol & 0.62  &  1.12  &  2.5  \\ 
Trehalose & 0.95 & 1.63 & 2.6    \\
Lysine hydrochloride & 0.90 & 1.38  & 2.7  \\
Arginine hydrochloride & 0.95 & 1.44 & 2.8  \\ 
\hline\hline
\end{tabular}
  \caption{The mass density, surface tension,
and the lowest vibrational frequency predicted by Eq.~(\ref{eq:nu_drop})
for drops of various liquids with a radius of $a = 50 nm$. The data for
Benzene and Diethylene glycol~\cite{CRC} are for droplets in air at 
room temperature.
The data for the three proteins~\cite{LT96} are for aqueous solutions at
approximately $50^\circ C$.  }
\end{table}

Recall that Eq.~(\ref{eq:nu_drop}) assumes a nonviscous liquid. In
fact, the viscosity of many of the liquids in Table 1 cannot be neglected
for such a small drop. When viscosity is sufficiently large, the drop
will not oscillate, but rather undergo overdamped motion~\cite{Chandra}.
The main lesson from the liquid drop model is that a {\it nonviscous} liquid
drop of $a = 50 nm$ with a typical surface tension and mass density would 
have a lowest vibrational frequency of the order of a few times $10^8 Hz$.

\section{An Elastic Sphere Model}

\begin{figure}
\begin{center}
\leavevmode\epsfysize=8cm\epsffile{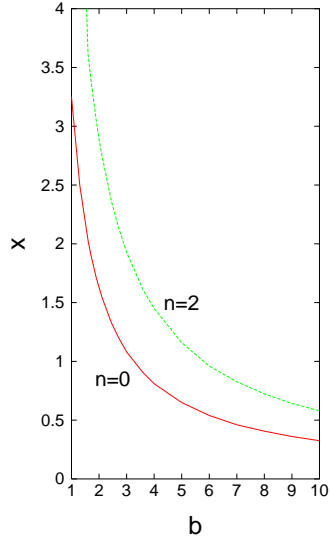}
\end{center}
\caption{The root $x$ of Eq.~(\ref{eq:eigenfreq}) is plotted as a function
of $b = {c_S}/{c_P}$ for $n=0$ (compression mode) and $n=2$ (quadrupole
deformation). }
\label{fig:bx}
\end{figure}

A better model for a virus particle is to treat it as a uniform elastic sphere.
The three independent parameters which characterize such a sphere can be
taken to be the radius $a$, the speed of pressure waves, $c_P$, and the
speed of shear waves, $c_S$. The oscillations of an elastic sphere are
treated in detail by Pao and Mow~\cite{PM}. Here we quote their results 
in the notation of Ref.~\cite{Bulatov}. The eigenfrequencies of the
normal modes are given by the vanishing of the determinant
\begin{equation}
\left| \begin{array}{ll}
              S_{rp}(n,x,y) & S_{rs}(n,y) \\
              S_{tp}(n,x)  & S_{ts}(n,y)
        \end{array} \right| = 0 \,,                \label{eq:eigenfreq}
\end{equation}
where 
\begin{eqnarray}
S_{rp}(n,x,y) = (n^2-n-\frac{1}{2} y^2)^2\, j_n(x) + 2 x j_{n+1}(x) \, , \\
S_{rs}(n,y) = n(n+1) [(n-1) j_n(y) - x j_{n+1}(y)] \, , \\
S_{tp}(n,x) = (n-1)j_n(x) - x j_{n+1}(x) \, , \\
S_{ts}(n,y) = -(n^2-n-\frac{1}{2} x^2)^2\,j_n(x) - x j_{n+1}(x) \, .
\end{eqnarray}
Here $x = \omega a/c_P$, $y = \omega a/c_S$, and the $j_n$ are spherical
Bessel functions. Let
\begin{equation}
b = \frac{c_S}{c_P}
\end{equation}
be the ratio of the speed of the shear wave to that of the pressure wave, so
that $y = b x$. Given $b$ and $n$, we can solve Eq.~(\ref{eq:eigenfreq})
for $x$ and hence for the frequency of the normal mode, $\nu = x\, c_P/(2 \pi a)$.
The smallest roots for $x$ for the $n=0$ and $n=2$ modes are plotted in
Fig.~\ref{fig:bx}. The corresponding frequencies of oscillation can be 
expressed as
\begin{equation}
\nu = 4.8 \times 10^9 Hz\, \left(\frac{50 nm}{a}\right)\,
\left(\frac{c_P}{1500 {m}/{s}}\right) \,  x \,.  \label{eq:nu_ball}
\end{equation} The frequencies obtained from this equation for the $n=0$ mode
are given in Table II for various materials.

\begin{table} 
\begin{tabular}{|c||c|c|c|} \hline\hline
Material  & $c_P$  & $c_S$  &  $\nu/(10^9 Hz)$ \\ \hline
Nylon~\cite{CRC2} & 2620  &  1070  &  11.4  \\
Polystyrene~\cite{CRC2}  &  2350  &  1120  &  11.3  \\ 
Polyethylene~\cite{CRC2}  &  1850  &  540  &  5.6  \\
Neoprene rubber~\cite{CRC2} & 1600 &  & 5 $x$  \\
Polynucleotides~\cite{Sar} & 1700 - 1900  &  & (5.3 - 6) $x$ \\
Amino acids~\cite{Sar}  & 1900 - 2400 &  & (6 - 7.7) $x$  \\
Globular proteins~\cite{Sar} & 1700 - 1800 &  & (5.3 - 5.7) $x$  \\
\hline\hline
\end{tabular}
  \caption{The speeds of sound (in m/s)
and the lowest vibrational frequency for the $n=0$ mode predicted by 
Eq.~(\ref{eq:nu_ball}) for spheres of various materials with a radius of 
$a = 50 nm$. In cases where no data for $c_S$ are available, the frequency
is given as a multiple of $x$, which is likely to be less than one for these 
materials. }
\end{table}

We can see that for a wide range of materials, the lowest mode of vibration
is a purely radial mode with a frequency of the order of a few times 
$10^9 Hz$ for a sphere of radius $a = 50 nm$.

\section{Conclusions}

In the previous sections, we have examined two models for a spherical virus 
particle, a liquid drop model and an elastic sphere model. It is of interest
that the two models yield estimates for the lowest vibrational frequency
which differ by only about one order of magnitude. Of the two models, the
elastic sphere model is probably the better description of a virus particle.
An even better model might be one in which the particle has a liquid core
(DNA or RNA) surrounded by an elastic outer shell (the capsid). Such a
model would probably yield vibrational frequencies intermediate between
those predicted by the two models discussed in this paper. In any case, 
we obtain an estimate for the lowest vibrational frequency of the same 
order of magnitude as that given in Ref.~\cite{Bab}, in the range of a few
GHz for particles with a size of about $100 nm$. Of course, the existence of
a resonance requires that damping be below the critical value above which
overdamped motion occurs. Even if this condition is fulfilled, it is difficult
to predict the width of the resonance. This remains a question for experimental
investigation. The existence of well defined resonances could prove valuable
both for basic science and for medicine. Thus this is a potentially fruitful
area for further research.
 
\begin{acknowledgments}
I would like to thank V. Ford and D. weaver for helpful comments.
  This work was supported in part by the National
Science Foundation under Grant PHY-9800965.
\end{acknowledgments}


\begin{thebibliography}{28} 

\bibitem{Bab} M. Babincov{\'a}, P. Sourivong and P. Babinec, Medical Hypotheses,
{\bf 55}, 450 (2000). 
   
\bibitem{Cooper} M.A. Cooper, F.N. Dultsev, T. Minson, V.P. Ostanin, C. Abell,
and D. Klenerman, Nature Biotechnology {\bf 19}, 833 (2001).

\bibitem{Cerf79} R. Cerf, B. Michels, J.A. Schulz, J. Witz, P. Pfeiffer
and L. Hirth, Proc. Natl. Acad. Sci. USA, {\bf 76}, 1780 (1979). 

\bibitem{Michels85} B. Michels, Y. Dormoy, R. Cerf and  J.A. Schulz,
J. Mol. Biol. {\bf 181}, 103 (1985).

\bibitem{WB01} J. Witz and F. Brown, Archives of Virology, {\bf 146}, 
2263 (2001).

\bibitem{Bulatov} V.L. Bulatov, R.W. Grimes and A.H. Harker, Phil. Mag. Lett.
{\bf 77}, 267 (1998).

\bibitem{Chandra} S. Chandrasekhar, {\it Hydrodynamic and Hydromagnetic
Stability}, (Oxford, 1961) Sect. 99.

\bibitem{CRC} D.R. Lide, ed., {\it CRC Handbook of Chemistry and Physics},
(CRC Press, Boca Raton, Florida, 2002), pp 3-26, 3-159, 6-149, 6-150.

\bibitem{LT96} T-Y Lin and S. N. Timasheff, Protein Science, {\bf 5}, 372 (1996). 

\bibitem{PM} Y-H Pao and C-C Mow, {\it Diffraction of Elastic Waves and Dynamic 
Stress Concentrations}, (Crane Russak, New York, 1973), Chap. 6, Sect. 2.

\bibitem{CRC2} D.R. Lide, ed., {\it CRC Handbook of Chemistry and Physics},
(CRC Press, Boca Raton, Florida, 2002), p14-41.

\bibitem{Sar} A.P. Sarvazyan, Annu. Rev. Biophys. Biophys. Chem.
{\bf 20} 321 (1991).



\end{thebibliography}
\end{document}